\documentclass[a4paper,fleqn]{cas-dc}
\usepackage[ruled]{algorithm2e}
\usepackage{bbding}
\usepackage{booktabs}
\usepackage{threeparttable}
\usepackage{tabularx}	
\usepackage{makecell}
\usepackage{graphicx}
\usepackage{float}
\usepackage{subfig}

\usepackage[numbers]{natbib}

\def\tsc#1{\csdef{#1}{\textsc{\lowercase{#1}}\xspace}}
\tsc{WGM}
\tsc{QE}

\begin{document}

	\let\WriteBookmarks\relax
	\def\floatpagepagefraction{1}
	\def\textpagefraction{.001}
	
	\title [mode = title]{Enhancing Spectral Knowledge Interrogation: A Reliable Retrieval-Augmented Generative Framework on Large Language Models}
	
	
	
	%

	\author[1]{Jiheng Liang}[orcid=0009-0004-5522-0884]
	\ead{liangjh65@mail2.sysu.edu.cn}
        \fnref{equal}
        \fntext[equal]{Jiheng Liang and Zujie Xie contribute equally to the article}
	\credit{Conceptualization, Investigation, Methodology, Validation, Data curation, Formal analysis, Software, Visualization, Writing}

	\author[1]{Zujie Xie}[orcid=0009-0000-1945-1045]
        \fnref{equal}
	\ead{xiezj8@mail2.sysu.edu.cn}
    
	\author[2]{Ziru Yu}[orcid=0000-0002-9763-8806]
	\ead{yuziru@mail.sysu.edu.cn}
	\cormark[2]
	\cortext[2]{Corresponding author. School of electronics and communication engineering, Sun Yat-Sen University, Shenzhen, 518107, China}

	\author[1]{Xiangyang Yu}[orcid=0000-0001-6673-3221]
	\ead{cesyxy@mail2.sysu.edu.cn}
	\cormark[1]
	\cortext[1]{Corresponding author. School of Physics,State Key Laboratory of Optoelectronic Materials and Technologies,Sun Yat-Sen University, Guangzhou, 510275, China}

	\affiliation[1]{
		organization={School of Physics, State Key Laboratory of Optoelectronic Materials and Technologies,Sun Yat-Sen University},
		city={Guangzhou},
		postcode={510275}, 
		country={China}
		}	
	\affiliation[2]{
		organization={School of electronics and communication engineering, Sun Yat-Sen University},
		city={Shenzhen},
		postcode={518107}, 
		country={China}
		}
	
	
	
	
	\begin{abstract}
	Large Language Model (LLM) has demonstrated significant success in a range of natural language processing (NLP) tasks within general domain. The emergence of LLM has introduced innovative methodologies across diverse fields, including the natural sciences. Researchers aim to implement automated, concurrent process driven by LLM to supplant conventional manual, repetitive and labor-intensive work. In the domain of spectral analysis and detection, it is imperative for researchers to autonomously acquire pertinent knowledge across various research objects, which encompasses the spectroscopic techniques and the chemometric methods that are employed in experiments and analysis. Paradoxically, despite the recognition of spectroscopic detection as an effective analytical method, the fundamental process of knowledge retrieval remains both time-intensive and repetitive. In response to this challenge, we first introduced the Spectral Detection and Analysis Based Paper(SDAAP) dataset, which is the first open-source textual knowledge dataset for spectral analysis and detection and contains annotated literature data as well as corresponding knowledge instruction data. Subsequently, we also designed an automated question-and-answer(Q\&A) framework based on the SDAAP dataset, which can retrieve relevant knowledge and generate high-quality responses by extracting entities in the input as retrieval parameters. It is worth noting that: within this framework, LLM is only used as a tool to provide generalizability, while RAG technique is used to accurately capture the source of the knowledge.This approach not only improves the quality of the generated responses, but also ensures the traceability of the knowledge. Experimental results show that our framework generates responses with more reliable expertise compared to the baseline.

	\end{abstract}
	\vspace{\topsep}

	\vspace{\topsep}
	\vspace{\topsep}
	\maketitle

\section{Introduction}
\hspace{1em}Since Chat-GPT\cite{openai2022} came out of nowhere in late 2022, the concept of Large Language Model (LLM), which is a sophisticated deep learning model based on the Transformer with parameter sizes reaching into the tens of billions, has come back to the forefront of researchers' minds. The development of LLM has got significant attention due to the extensive knowledge and impressive interaction with humans. What's more, the incredible ability of LLM about extracting implicit information from prompts with appropriate instruction-following distinguishes itself among most of previous deep learning models. Compared to their previous smaller counterparts, LLMs also demonstrate potent generalisation across various Natural Language Processing (NLP) tasks, illustrating their capacity to resolve unseen or intricate challenges in different domains. Currently, LLMs have demonstrated commendable performance in general domains\cite{le2023bloom}\cite{dubois2024alpacafarm}\cite{du2021glm}\cite{wang2018glue}. Furthermore and naturally, researchers want to introduce LLM in the natural science, a field that places a high demand on logical thinking, to relieve the extremely time-consuming and labour-intensive in practical applications. For instance, spectroscopy-based detection technology is a widely used analytical method with important applications in both the natural sciences and industry. However, for arbitrary samples, the spectroscopic techniques (e.g., Ultraviolet spectrum; Near-Infrared spectrum) and stoichiometric methods (e.g., Preprocessing method; Machine learning method) used in the experiment need to be determined based on past relevant studies because of the poor migration of model. Researchers have to spend a lot of time on the information collection of relevant data in the preliminary stage of the research, which is the most time-consuming and repetitive task in spectral analysis. Now, LLM can learn from large knowledge database and then can be used to provide related information about analyzed object including different points mentioned earlier, which may fully accelerate the procedure of spectral detection in different objects.

\hspace{1em}However, a significant challenge in adapting LLM to the domain of spectral detection is the phenomenon of hallucination\cite{ji2023towards} concerning specialized knowledge. The general knowledge possessed by these models frequently proves inadequate when applied to specialized fields, primarily due to a deficiency in domain-specific expertise. Figure 1 shows the limitations of Chat-GPT 4 in answering questions in the field of spectral detection without source. Without the support of professional knowledge, it is difficult for general large models such as Chat-GPT 4 to generate accurate answers in professional fields. Given the high demand for reliability in the field of spectral detection, it is essential to enhance the knowledge base associated with LLM. This initiative is critical to ensure that LLM can provide accurate and dependable information to researchers engaged in spectral analysis methodologies. Instruction-tuned LLM, exemplified by InstructGPT\cite{ouyang2022training}, serve as the primary method to mitigate LLM into novel domains. This adaptation is achieved by constructing Instruction Fine-Tuning (IFT) data from knowledge database, thereby augmenting the expertise of LLM through supervised fine-tuning (SFT) and improving its interactive capabilities.
\begin{figure*}[h]
  \centering
  \includegraphics[width=0.9\textwidth]{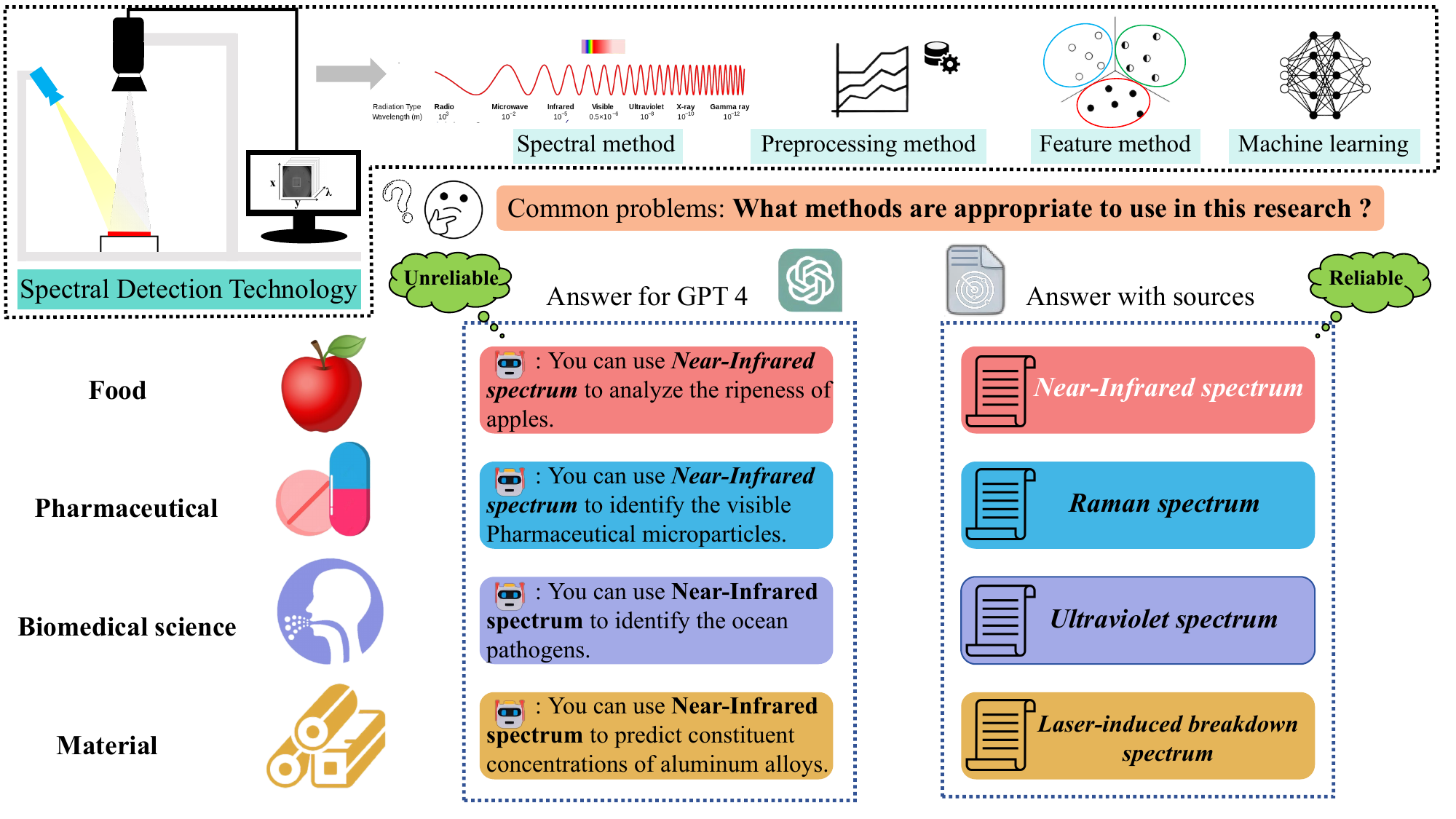}
  \caption{The limitations of GPT 4 to answer questions in the field of spectral detection without sources}
  \label{fig:example}
\end{figure*}

\hspace{1em}But Initially, it is important to note that the majority of existing relevant datasets are predominantly concentrated in the domains of bioscience\cite{bhattacharya2024large} and medicine\cite{thirunavukarasu2023large}. In contrast, the field of spectral analysis, which is considered relatively traditional, is characterized by a limited availability of open-source datasets. Most of datasets in this domain is primarily represented by spectral curves of specific sample types and is notably deficient in textual datasets suitable for NLP applications. Consequently, if there is an intention to adapt LLM to a spectral-related domain, it may be necessary to independently develop the requisite scientific datasets.Furthermore, it is indeed more challenging than one might expect for LLM can comprehend specialized knowledge and generate accurate, knowledge-consistent responses solely through SFT method\cite{li2024getting} without undergoing additional pretraining. Recent studies have explored the integration of LLM with external application programming interfaces (APIs) to specific domains to improve the precision of model outputs\cite{gu2023apicom}\cite{thoppilan2022lamda}. However, these methodologies necessitate operation on external servers, and the associated tasks are both financially burdensome and subject to network limitations, which may impede the progress of LLMs in scientific advancement.Lastly, from the perspective of a natural science researcher, scholars frequently concentrate on the foundational sources of knowledge, such as specific literature or knowledge bases, to facilitate their subsequent exploration of additional relevant information. However, an approach that relies solely on Instruction-tuned for LLM refinement may necessitate further annotation of knowledge sources within the IFT data if the intention is to cite the source of knowledge in the generated responses. This method does not instill a high degree of confidence in the accuracy of the outputs, representing a significant limitation. An alternative strategy involves the use of Retrieval Augmented Generation (RAG)\cite{zhao2024retrieval}, which leverages specialized databases to retrieve pertinent knowledge from data sources and generate responses accordingly. In essence, RAG integrates information retrieval methodologies with the generative capabilities of LLM, thereby addressing the limitations of Instruction-tuned techniques by direct access to data sources through annotations within databases conveniently. Much previous work\cite{kanakarajan2021bioelectra}\cite{rasmy2021med} has used a similar approach to source knowledge.

\hspace{1em}In our work, considering that professional literature constitutes a valuable repository of advanced knowledge, research findings, and engineering methodologies, we firstly present the Spectral Detection and Analysis Based Paper (SDAAP) dataset as a foundational source of knowledge. SDAAP encompasses information from relevant publications spanning the years 2014 to 2023, with each entry meticulously categorized such as the research object, the spectroscopic techniques employed, and the associated chemometric parameters. In addition to the labeled literature, SDAAP incorporates IFT data derived from the insights of all the publications in dataset; each IFT data also includes the relevant knowledge and its corresponding literature source, facilitating SFT process. The total number of IFT data amounts to more than 20,000 entries. Subsequently, based on SDAAP, we developed a automatic Q\&A framework in related domain, which can parse and extract the entity and question formats present in a query, employing the parsing outcomes as query parameters to retrieve pertinent spectral detection knowledge. LLM can then reference this retrieved knowledge to generate a response to the input query. Furthermore, our framework does not rely on instruction-tuning to facilitate the acquisition of new knowledge by LLM, but rather uses it as tool to provide generalizability. Instead, pertinent knowledge can be obtained in a relatively controlled manner through retrieval techniques, thereby enhancing the quality and reliability of the generated responses and ensuring the provision of accurate knowledge sources. In summary, our contributions can be summarized as follows:

\hspace{1em}\textbullet We develop the SDAAP dataset, the first systematically organized open-source textual knowledge dataset for spectral analysis and detection. This dataset comprises annotated literature data alongside corresponding knowledge instruction data, thereby addressing a significant gap in textual datasets within the domain and establishing a foundational resource for the subsequent application of LLM in this area.

\hspace{1em}\textbullet We designed a knowledge quiz framework based on the SDAAP dataset. The framework can generates high-quality and reliable responses by parsing questions and retrieving the knowledge associated with them, thereby responding the question of various Q\&A scenarios within the domain of spectral detection analysis and reducing the repetitive labour. Further details regarding the project are accessible on our GitHub page: coming soon.

\hspace{1em}\textbullet Within our question-and-answer framework, we integrate techniques of Instruction tuning and retrieval-augmented generation (RAG). The LLM serves only as a tool to enhance generalizability, while RAG techniques are employed to accurately acquire the source of knowledge, thereby ensuring traceability of knowledge While augmenting quality of response.

\section{Datasets}
\hspace{1em}As is mentioned previously, the process of retrieving pertinent information within the context of industrial grading and detection through spectral analysis is predominantly characterized by repetitive works, resulting significant inefficient use of time. Leveraging LLM, an advanced computer science technology, can effectively streamline these repetitive tasks and release a substantial amount of human resources. In order do this, our research endeavors to incorporate LLM into spectral analysis to provide rapid and dependable responses to queries based on existing knowledge. 

\hspace{1em}Recently, the mainstream approach to migrating LLM to a new domain has been to fine-tune LLM based on specialized corpora so that it can be adapted to the corresponding verticals\cite{xie2023darwin}. However, very few existing open-source corpus datasets related to scientific and technical disciplines contain expertise related to spectral analysis. This means that there is a lack of professional datasets tailored for spectral analysis. Given the limited availability of open-source corpus dataset related to spectral analysis, this study opts to independently create a relevant dataset and its corresponding corpus from literature within the professional field. These corpus resources are then applied in our designed framework (see in Section 3) for reliable spectral detection knowledge question and answer (Q\&A) tasks. This section outlines the methodology employed for dataset construction, as well as the distribution of labels sourced from the dataset literature. Subsequently, a corpus dataset is constructed and made available, which can be utilized for various future research endeavors.

\subsection{Paper collection}
\hspace{1em}In the context of paper collection focusing on spectral analysis using machine learning methods, the Web of Science was employed as an indexing tool to gather relevant scholarly literature comprising various domains (e.g. food, biology, energy…). Considering the prevalent use of machine learning methods in interdisciplinary research since 2015 approximately, scholarly paper published within the past decade (2013-2023) were selected for analysis. Various keyword combinations were utilized for the search, and subsequent operations, including paper de-duplication by human intervention, were conducted on the aggregated search results. A total of 4461 thesis were obtained through this screening process. It is important to note that all these papers are accessible in full-text format from reputable publishers like Nature, Springer, Elsevier, MDPI, among others. The scope of this resource transcends the English language to encompass a broader spectral detection of related knowledge. In the following Figure 2, an analysis of the publication timeline of these papers reveals a noticeable increase in the application of spectral analysis with machine learning method in detection and other areas. This trend underscores the significance of the research endeavor. Subsequently, a web-scraping tool was utilized to extract content from various publishers and convert it into plain text for further processing.

\begin{figure*}[h]
  \centering
  \includegraphics[width=1.1\textwidth]{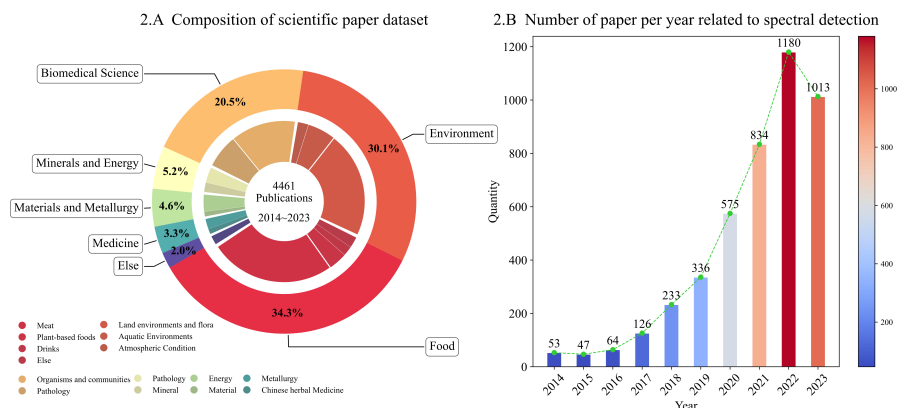}
  \caption{Distribution of scientific paper dataset}
  \label{fig:example}
\end{figure*}

\subsection{Labelling and Indexing}
\hspace{1em}During a comprehensive spectral analysis investigation, researchers consistently focus on specific configurations at first. To effectively retrieve essential information from academic literature, it is essential to categorize profession academic papers into distinct labels that can be seamlessly integrated into our spectral Q\&A framework, thereby facilitating further research by independent scholars.

\hspace{1em}In the first stage of spectral detection studies, researchers must select specific spectral methods, such as Near-Infrared spectrum (NIR), Ultraviolet spectrum (UV), Raman spectroscopy, among others, depending on the characteristics of the object and properties under investigation. Traditionally, this kind of process has been time-consuming as researchers are required to repeatedly search for and review numerous papers that are relevant to their study topic. Consequently, we primarily categorize this type of information across all papers within our datasets into the special kind of Label A. Through the automated retrieval facilitated based on Label A by our framework, researchers are able to efficiently access pertinent information related to their own research with different spectral detection method, thereby streamlining the process and minimizing tedious tasks.

\hspace{1em}Once the spectral method employed in the experiment and its pertinent specifics have been identified, the selection of the machine learning technique and its associated parameters becomes crucial during the data processing stage. We synthesized and categorized the machine learning information extracted from the papers in our datasets into a class of Label B that is distinct from Label A, including preprocessing techniques, feature processing methods, and models for further analysis.

\hspace{1em}In brief, the extracted labels from any one of papers in our datasets can be divided into two primary sections: Label A is utilizes for summarizing essential information, such as spectral method, and directing researchers towards papers that are most pertinent to their inquiry; Label B is focuses on providing insights into machine learning techniques employed in the paper. These labels aid in pinpointing relevant and valuable literature that pertains to various inquiries posed in the knowledge quiz concerning spectral detection. 

\hspace{1em}For instance, we presented a selection of labels derived from some papers within our datasets in the Figure 3 provided below. The methodology involves utilizing Chat-GPT to extract labels from the papers, followed by a manual data cleaning process to rectify any inaccuracies. Notably, majority of labels are only obtained from the abstract section of the papers. Since that abstracts encapsulate the core elements of the literature, encompassing objectives, methods, conclusions, etc., most of labels we need can be extracted from it, and this minimizes copyright issues. In instances where labels cannot be directly derived from the abstract, such as preprocessing methods and machine learning models, the paper is converted into embedding vectors. Subsequently, relevant answers are retrieved based on maximum cosine similarity to extract the corresponding labels.

\begin{figure*}[h]
  \centering
  \includegraphics[scale=0.5]{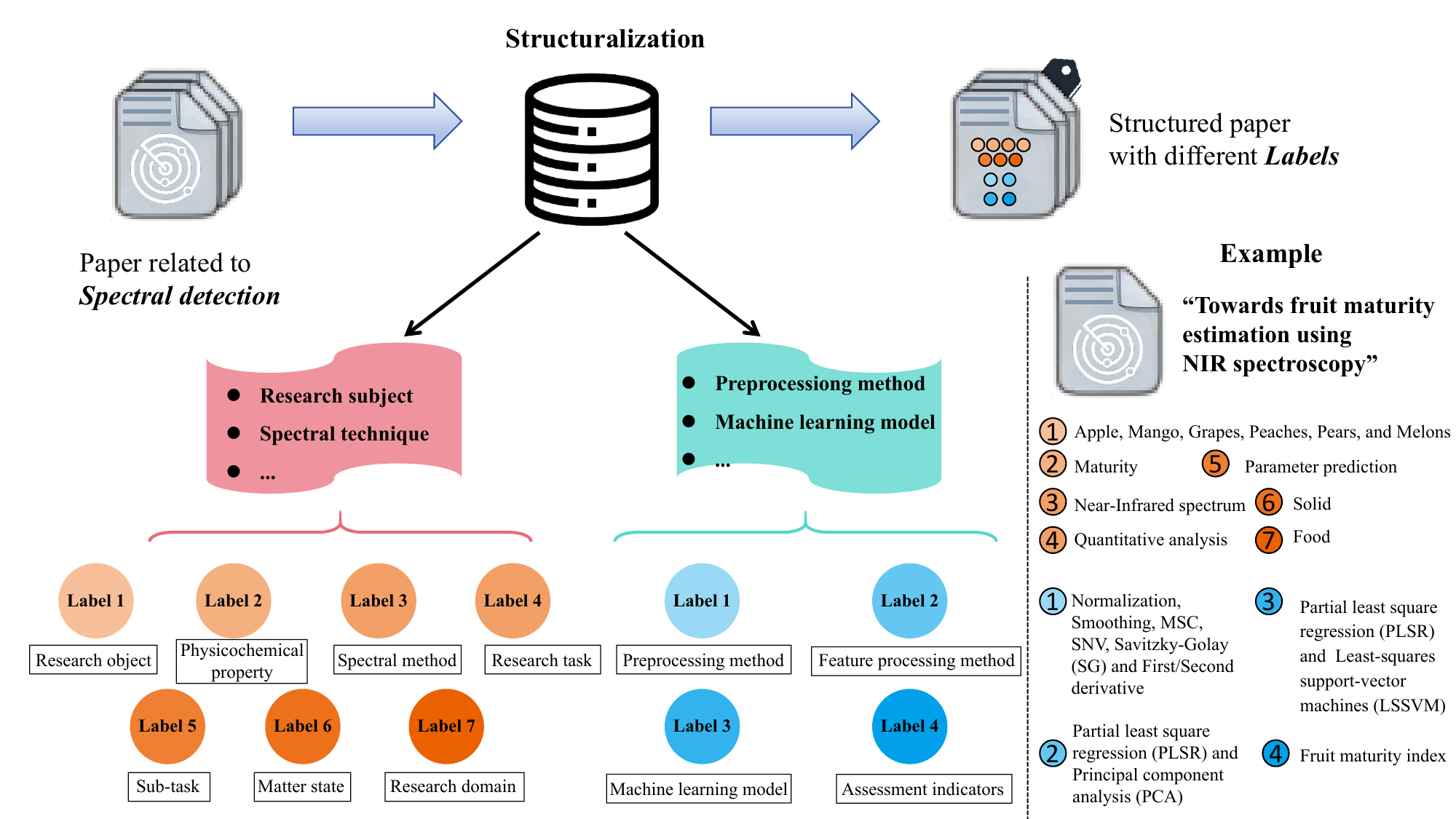}
  \caption{Two categories of labels of scientific paper}
  \label{fig:example}
\end{figure*}

\subsection{Q\&A corpus data based on spectral knowledge}
\hspace{1em}It is widely acknowledged that Instruction Fine Tuning (IFT) data is crucial for the implementation of LLM in vertical domain and other Natural Language Processing tasks. Due to the absence of specialized IFT data for spectral detection and analysis, we have developed a framework to create IFT data automatically for each labeled literature in our datasets using Chat-GPT. This framework consists of four distinct stpdf as follows.

\subsubsection{Question focus}
\hspace{1em}Drawing upon prior knowledge within specific fields, we initially selected commonly encountered questions in practical scenarios. As an example, in the context of spectral analysis and detection research, we focus the category of spectral method, preprocessing methods, feature processing methods, metrics and outcomes, as well as machine learning models. These topics are frequently addressed in spectral analysis studies and are utilized to compile our Q\&A corpus data.
\subsubsection{Question split and construction}
\hspace{1em}A template was developed to facilitate the generation of the question component of the corpus. This template consists of two parts: Part A, which serves to define the research object, and Part B, which is utilized to create various question. By inputting the appropriate label of the paper as a prompt, users can efficiently employ Chat-GPT to generate diverse formats of question. This template is designed to be easily adaptable to different research subjects. The structure of the template is illustrated in the Figure 4 below.
\subsubsection{Answer generation}
\hspace{1em}In the same way as second step, we embed the question and the label corresponding to the answer from paper into the prompt, and ask Chat-GPT to generate a formatted answer based on the information provided. Here, we only constructed limited types of Q\&A corpus data mentioned in section 2.3.1. To acquire additional Q\&A data, researchers can simply substitute Part B questions corresponding to different labels.
\subsubsection{Data Cleaning}
\hspace{1em}While the utilization of Chat-GPT for generating IFT data has become prevalent in cutting-edge research, there are still challenges that require human intervention. For instance, the extensive automated generation of Q\&A data using LLM driven approach always encounter obstacles related to inappropriate representation, just like "This study used (Method A) to (Object B)". Actually, the utilization of the phrase "This study" as the subject in typical question and answer interactions is not recommended. If a substantial portion of the generated data exhibits such issues, it may affect the application of these corpus data in downstream tasks. To address these concerns, manual intervention is employed to rectify similar problems, for instance, by replacing phrases like " This study used (Method A) to (Object B)" with "Related studies show that (Method A) can be used in (Object B)".

\hspace{1em}Over all the following stpdf, we finally got high quality IFT data, totally including 22305 items based on different paper and each item include one question and matched answer with related knowledge in particular paper. All the corpus data were contained in our dataset and are publicly accessible.

\begin{figure*}[h]
  \centering
  \includegraphics[width=1.1\textwidth]{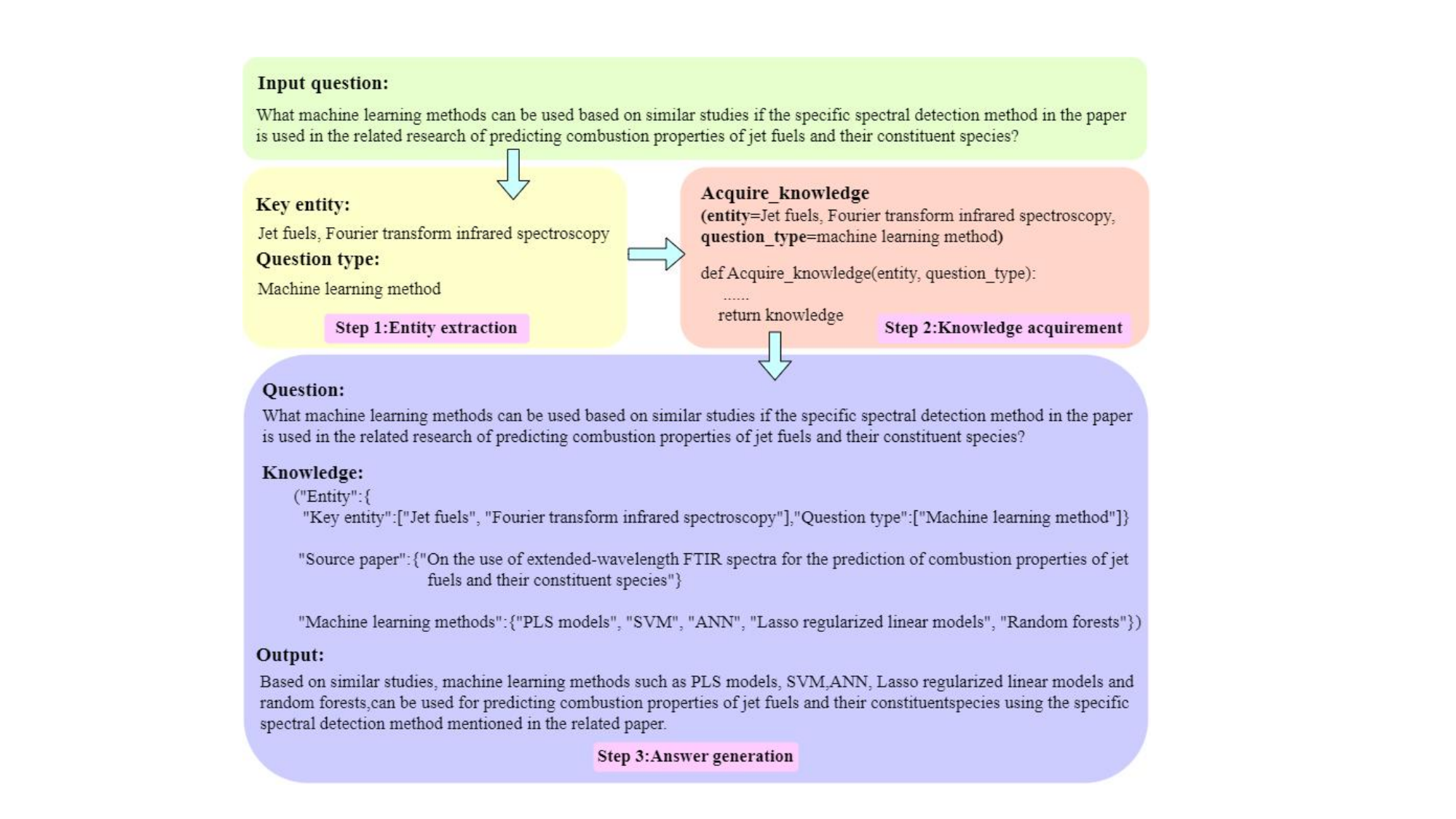}
  \caption{Process for knowledge-based response generation. Step 1:
  Extract the entity of the input question. Step 2: Acquire the knowledge from the SDAAP dataset. Step 3: Generate a response with acquiredknowledge.}
  \label{fig:example}
\end{figure*}

\section{Method}
\hspace{1em}As mentioned earlier, extensive language models have a broad spectrum of applications across various fields, with one prominent example being the utilization of large models to develop specialized domain question-and-answer (Q\&A) systems\cite{wang2023huatuo}. By employing fine-tuning techniques, these large models can be further customized using specific domain data to enhance their understanding of vertical domain-specific information. This method significantly enhances the performance of large language models in a multitude of contexts\cite{du2023calla}. The refined macro-models serve as modern expert systems, catering to the growing demand for tools that facilitate the rapid and accurate acquisition of pertinent knowledge to optimize time management. Nevertheless, the results produced by fine-tuned macro-language models may not always be dependable in domains where high-confidence outputs are essential, such as in the fields of biology, medicine, and industrial inspection. Furthermore, users often face challenges in accurately tracing the sources of the information provided in the responses, a particularly crucial aspect in scientific and technical research.

\hspace{1em}To address these challenges, we employed a dual strategy involving the refinement of instructional methods and the utilization of RAG (Retrieval-Augmented Generation) to develop a question-and-answer (Q\&A) framework centered around a comprehensive language model, as shown in the Figure 5 below. Leveraging a meticulously curated dataset enriched with specialized literature as a foundational knowledge repository, we operationalized the Q\&A system within the domain of spectral detection and analysis. The framework, powered by an advanced language model, is adept at processing diverse question formats ($q_i$), retrieving relevant information from the knowledge base, and delivering accurate responses ($r_i$). Given that our knowledge repository is structured around annotated literature sources, each piece of information can be directly linked back to its respective source, enabling transparent tracking of the expertise underpinning the generated responses. Subsequently, we outline the three key elements comprising the framework: entity extraction and question parsing, knowledge retrieval, and response generation.

\begin{figure*}[h]
  \centering
  \includegraphics[width=0.9\textwidth]{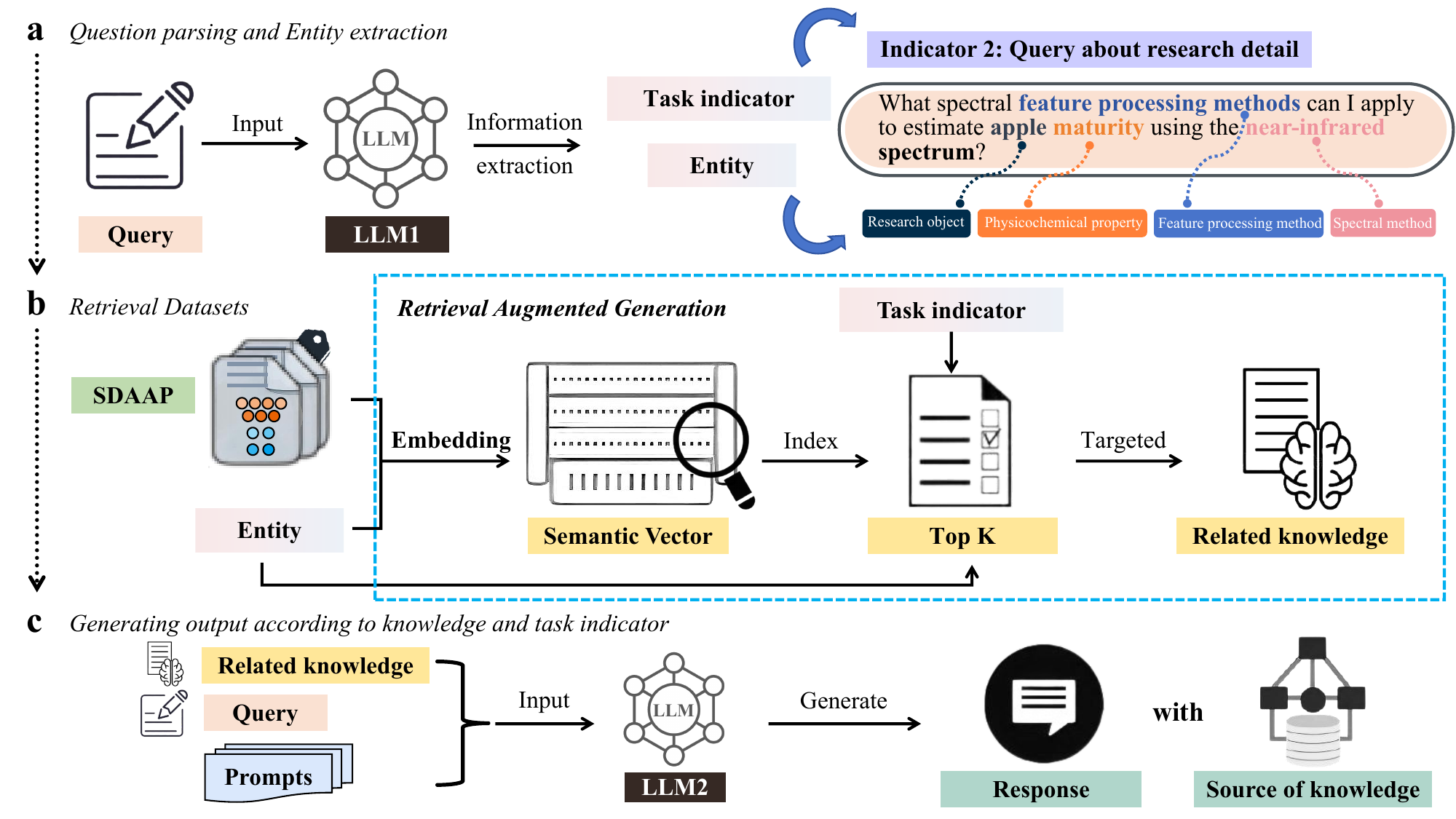}
  \caption{Detailed framework of the Q\&A system}
  \label{fig:example}
\end{figure*}

\subsection{Question parsing and Entity extraction}
\hspace{1em}For arbitrary question input, we need to obtain the object of related study indicated in the question at first. As an example, considering “In the related study on the prediction of sweetness in apples, …”, the process of extracting entity information related to the inquiry involves identifying the subject of the study (apples) and the specific aspect under investigation (sweetness). Typically, questions contain descriptive elements that pertain to the object of study, facilitating the extraction of relevant content in a more generalized manner.
\begin{align}
&M(Prompt_e,Ques) = E_{pre} \\
&Loss_{Entity}(Ques) = Loss_{Entity}(E_{pre},Groundtruth)
\end{align}
\hspace{1em}Furthermore, it is essential to extract information concerning entities that are pertinent to the orientation of the question. In the context of spectral detection, questions of interest to researchers can be categorized into two main groups: the first pertains to the selection of spectral detection methods in experiments, while the second involves the modeling and calculation procedures post data acquisition. These distinct types of questions necessitate different approaches for addressing them effectively. For questions falling under the first category, which may involve various spectral detection methods for a given study, it is crucial to identify and compare analytical methods from relevant literature, considering factors such as accuracy and experimental conditions. Conversely, questions related to the second category may require additional information on the spectral category used and the specific objective of the inquiry to facilitate knowledge retrieval and response generation.

\hspace{1em}To differentiate between these two question categories, a dichotomous indicator termed "$task\_indicator$" is introduced in the questioning process. This indicator helps to categorize questions and determine the need for acquiring additional entity information based on the nature of the inquiry. For the first category of questions, there is no need to obtain any additional entity information; for the second category of questions, in addition to the previously obtained relevant content of the research object, we need to additionally obtain the two parts of the adopted spectral category and the question objective for subsequent knowledge retrieval and response generation. Given its capacity for generalization across various input forms and adeptness at addressing diverse inquiries, LLM-1 was employed in this study to extract entity information of two distinct types. Through fine-tuning the LLM-1 with task-specific data, it was able to efficiently and precisely retrieve the specified entity information. Further elaboration on the experimental results can be found in Section IV.
\begin{figure*}[h]
  \centering
  \includegraphics[width=0.9\textwidth]{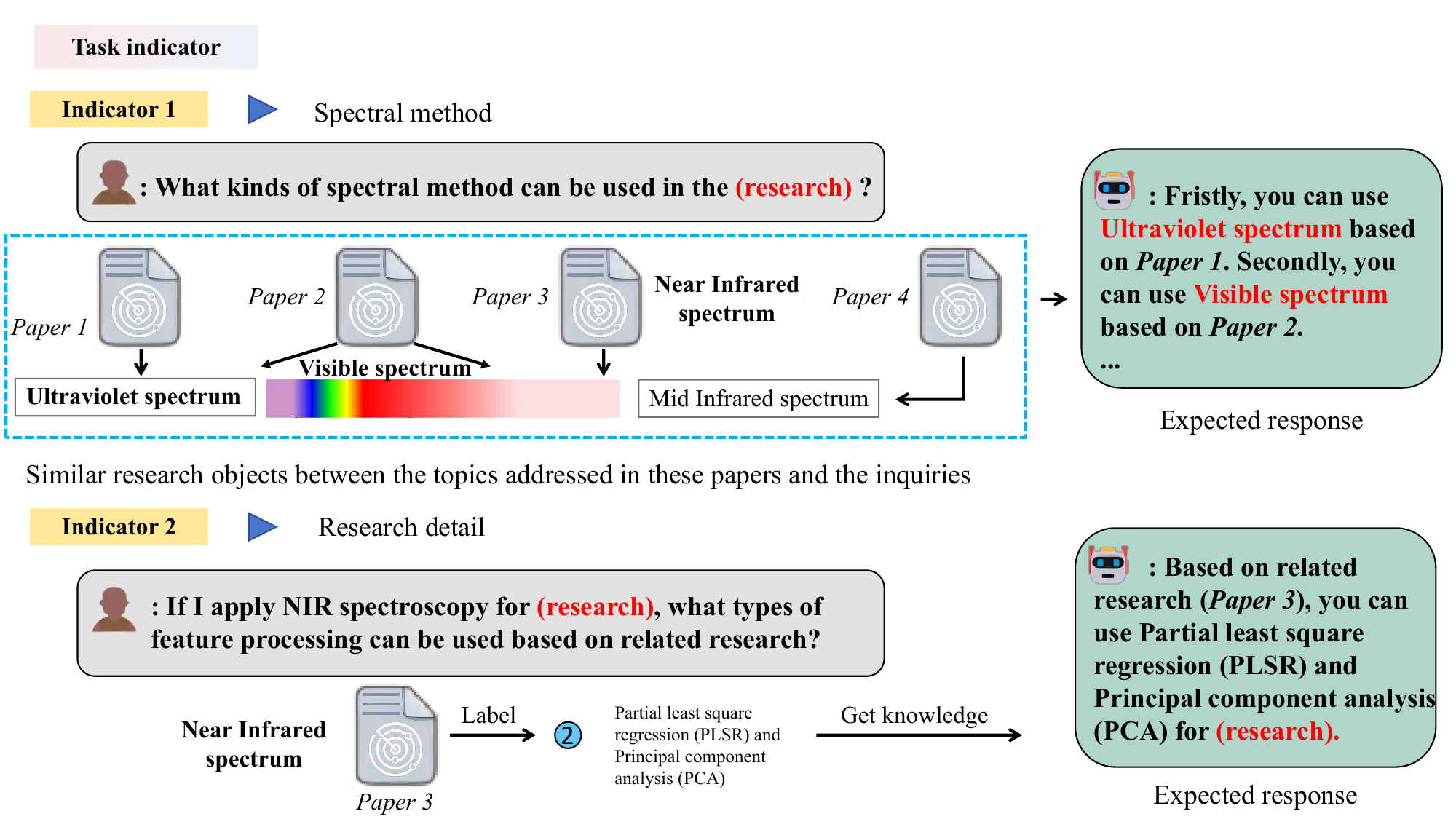}
  \caption{Different approaches to two types of questions}
  \label{fig:example}
\end{figure*}

\subsection{Retrieval Datasets}
\hspace{1em}Following the extraction of entities, the query is transformed into a tuple comprising various instances, denoted as ($q_i$, $e_i$, $task\_indicator$), which includes the original query, the identified entities, and the task indicator. During the process of knowledge retrieval, the parameter e\_i is utilized to retrieve information from the knowledge repository. By organizing each document in the knowledge repository and extracting the relevant text labels, the challenge of document retrieval can be reframed as determining the similarity between the entity details extracted from the query and the document labels. Consequently, for knowledge retrieval, a cosine similarity retrieval approach based on vector embedding is employed. This involves evaluating the resemblance between the entities extracted from the query and the corresponding labels in the literature to identify highly pertinent literature related to the query subject. In addition to cosine similarity computation, two other methods were chosen as baseline techniques for comparison: BM25, a statistically driven retrieval method\cite{kadhim2019term}; and the bag-of-words (BoW) model\cite{zhang2010understanding}. By contrasting the performance of cosine similarity retrieval with these two baseline methods within the knowledge repository, the effectiveness and precision of each method can be evaluated. Detailed experimental data and outcomes are presented in Section IV.
\begin{align}
&M(Prompt_K,Task_indicator,Knowledges) = knowledge
\end{align}
\hspace{1em}Moreover, following the identification of relevant literature using cosine similarity retrieval, diverse knowledge extracted from the literature label based on the task\_indicator is utilized to address the specific query associated with corresponding label. For instance, in cases where the task\_indicator is denoted as the first category of questions, the top 10 literature pieces with the highest cosine scores are selected to present various spectral analysis techniques applicable for examining the research subject. Conversely, when the task\_indicator is identified as the second category of questions, pertinent information is derived from the structured labels of the literature, such as preprocessing methodologies, feature processing techniques, and the machine learning models employed. In the other hand, where the queried object in the task\_indicator is not predefined within the labels, the entire Abstract section of literature is considered as the source of knowledge. Although the abstract may not contain specific information pertinent to the inquiry at hand, it serves as a valuable repository of literature that could potentially offer pertinent insights for the research.

\subsection{Generating output according to knowledge and task indicator}
\hspace{1em}Finally, based on the original input query $q$, the retrieved expertise $k$, and the template prompt $Prompt_k$ that splices the above two, we fine-tune a new LLM for generating the response $R$, viz:
\begin{align}
&M(Prompt_K,q,c) = R_{pre} \\
&M(Prompt_A,q)=A_{pre} \\
    &Loss_{Attribute}(q)=Loss_{Attribute}(A_{pre},Groundtruth)
\end{align}
\hspace{1em}Here, since we have already acquired knowledge matching the question asked as part of the input through the retrieval method, we do not need to infuse the knowledge into the big model by way of fine-tuning, thus circumventing the inherent shortcomings of inaccurate generation and untraceable sources that may result from acquiring knowledge directly from the big model. Our aim here in using LLM to generate responses is to ensure the diversity\cite{wang2023knowledge}, specialization, and fluency of the generated responses; in other words, we want to generate responses with a tone style that is closer to that of real researchers, which is the part that fine-tuning techniques excel at, as they can generate responses of similar style with a small amount of manually processed data. Therefore, we first randomly extracted some knowledge from the dataset and constructed it into instances ($Prompt_k$, $q$, $c$) through a framework, and then generated responses using the OpenAI API\cite{openai2022}. We then recruited three annotators with research backgrounds in spectral analysis who manually adjusted the OpenAI-generated responses to more closely resemble the descriptive style of professionals, and we collated all manually processed responses for supervised fine-tuning of the LLM. We used both NLP metrics and AI to evaluate the performance of the fine-tuned model in generating responses and compared it with other baseline models, and the experimental results can be found in Section IV.
\begin{figure*}[h]
  \centering
  \includegraphics[width=0.9\textwidth]{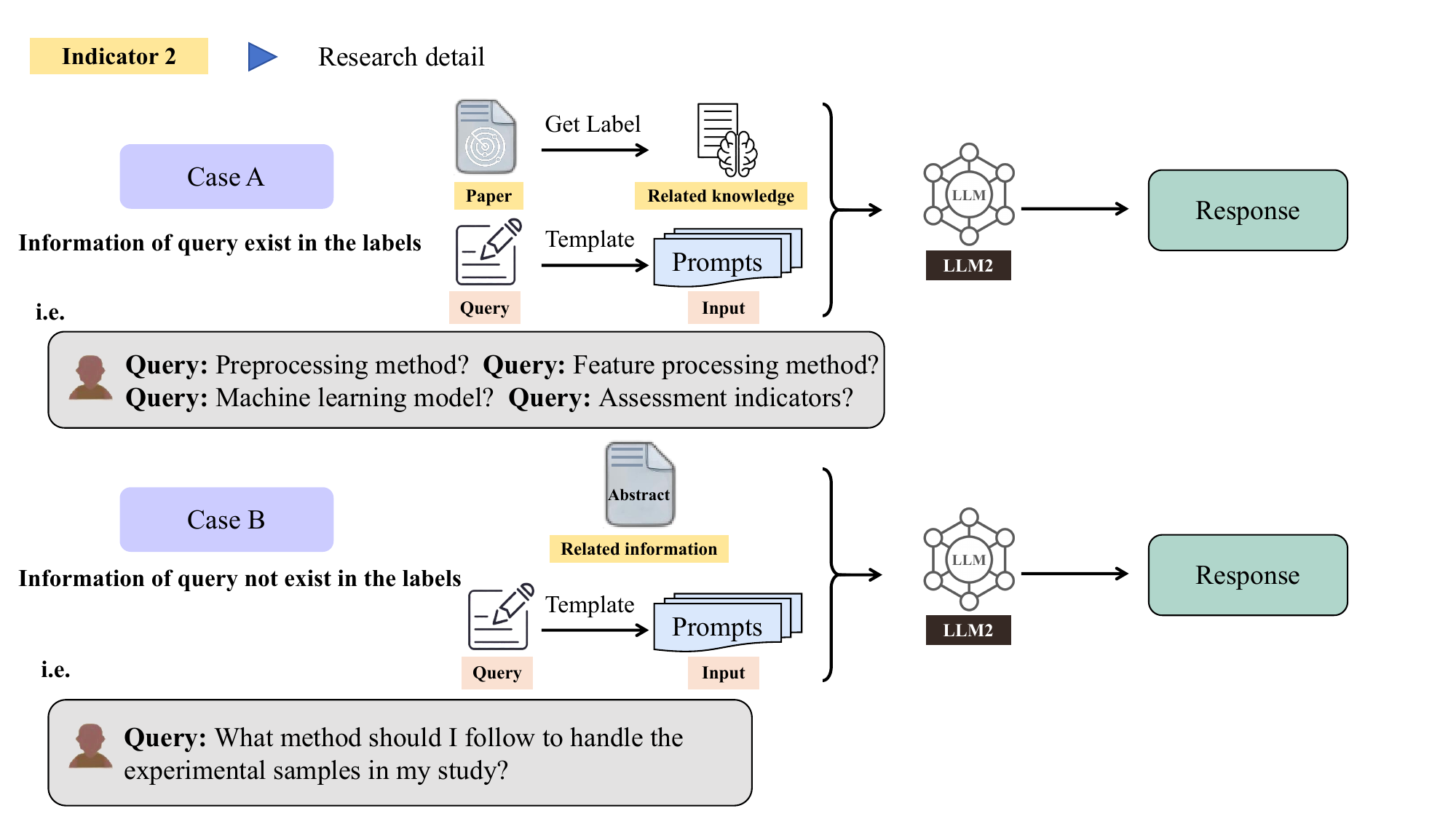}
  \caption{Different approaches to two cases whether information of query exist in the labels}
  \label{fig:example}
\end{figure*}

\section{Experiment}
\subsection{Detail of Implementation}
\hspace{1em}Our experimental setup aims to effectively evaluate the performance of our proposed method in the field of searching for papers related to spectral detection. All experiments were performed on a single machine equipped with an NVIDIA RTX 4090 GPU and 24GB of memory to ensure efficient execution and minimal processing time. The experiment is mainly divided into three parts. The first step is to use the fine-tuned Llama2-7b model to extract entities from the input question and the second step is to use the TF-IDF cosine similarity retrieval method to retrieve the corresponding literature and knowledge in the paper database based on the extracted entity keywords. At last, the fine-tuned Llama3-8b model was used to generate answers based on the literature knowledge.

\hspace{1em}In our experimental setup, we utilized the Low-Rank Adaptation (LoRA) technique to fine-tune both the Llama2-7b and Llama3-8b models, which is a parameter-efficient fine-tuning method that introduces trainable low-rank matrices into each layer of a pre-trained model, significantly reducing the number of trainable parameters and making the fine-tuning process more efficient and less resource-intensive\cite{hu2021lora}. The approach allows for the adaptation of large pre-trained models to specific tasks while preserving their performance. The hyperparameters set in the experiment is shown in Table 1.
\begin{table}[!ht]
\caption{Hyperparameters in experiment}
\label{tab:Hyperparameters in experiment}
    \renewcommand{\arraystretch}{1.3}
    \centering
    \scalebox{0.9}{
    \begin{tabular}{ll}
    \hline
        Hyperparameters & Values \\ \hline
        Batch size & 128 \\
        Max epoch & 15 \\
        Learning rate  & 3e-4 \\
        LoRA rank & 16 \\
        LoRA alpha & 32 \\
        LoRA dropout & 0.05 \\
        LoRA target modules & q\_proj, v\_proj \\ \hline
    \end{tabular}}
\end{table}

\hspace{1em}By employing these parameters, we ensured that the LoRA fine-tuning process was both effective and efficient. This enabled the Llama2-7b model to accurately extract entities from input questions and the Llama3-8b model to generate high-quality answers based on the retrieved literature and knowledge. Consequently, we leveraged the strengths of large pre-trained models while maintaining computational feasibility within our hardware constraints, facilitating an effective evaluation of our proposed method in the domain of spectral detection.
\subsection{Base models and Baseline}
\hspace{1em}For entity extraction from the input question, the Llama2-7b model fine-tuned by LoRA is utilized and called LLM1, which is an advanced language model known for its enhanced contextual understanding and language generation capabilities\cite{touvron2023llama}. The task of LLM1 is to identify key entities related to spectral detection and focus the retrieval process on the most relevant aspects. The extracted entities include: the research object in the input question, the spectral method mentioned in the input question, and the question type of the input question. After entity extraction, we searched our database using a cosine similarity-based retrieval method which converts the text into TF-IDF feature vectors and calculates the cosine similarity between the extracted entity vectors and the entity vectors present in the research papers\cite{qaiser2018text}, allowing us to retrieve the papers with the highest similarity scores and ensuring that our retrieval process is both targeted and relevant.

\hspace{1em}Two models were used respectively to generate detailed responses: Llama2-13b fine-tuned on LoRA and Llama13-8b fine-tuned on LoRA, collectively referred to as LLM2. Llama3, particularly the Llama13-8b model, represents a further advancement in language modeling, offering superior performance in both understanding and generating natural language\cite{meta2024introducing}. During the process, the top five retrieved papers are fed into LLM2, which synthesizes relevant content to provide a comprehensive answer to the input question. This step leverages the advanced language generation capabilities of LLM2 to ensure high-quality and contextually accurate responses.

\hspace{1em}The baseline model we used for the experiments is Chat-GPT 3.5 owing to its cost-effectiveness rather than GPT 4.0, which is a mature model known for its powerful language generation capabilities\cite{west2023ai}. Chat-GPT 3.5 is pre-trained on a diverse dataset and performs well in natural language understanding and generation tasks, allowing us to benchmark the performance of our system in terms of entity extraction, retrieval accuracy, and response generation quality. The comparative analysis between Chat-GPT 3.5 and our Llama2-based model highlights the progress we have made with our focused approach in the area of searching for papers related to spectral detection.
\subsection{Metrics}
\hspace{1em}In our experiments, we adopted a comprehensive set of evaluation metrics to evaluate the performance of the models in entity extraction (LLM1) and response generation (LLM2).

\hspace{1em}For the entity extraction task using LLM1, the main evaluation metrics include BLEU\cite{papineni2002bleu}, ROUGE\cite{lin2004rouge}, METEOR\cite{banerjee2005meteor}, BERTScore\cite{zhang2019bertscore} and accuracy (Acc). BLEU, ROUGE, and METEOR are traditional metrics for evaluating text quality by comparing generated text with reference text. BERTScore leverages BERT's contextual embeddings to provide a more nuanced measure of similarity between generated and reference text. Accuracy (Acc) measures the overlap between the entities extracted by the model and the reference entities, providing a direct indicator of the correctness of entity extraction.

\hspace{1em}For the response generation task using LLM2, we used similar metrics: BLEU, ROUGE, METEOR, BERTScore, and AI evaluation. BLEU, ROUGE, and METEOR evaluate the generated responses against the reference answers while BERTScore provides a more sophisticated semantic similarity measure. In addition, Chat-GPT4 is used for AI evaluation, which evaluates responses against specific criteria shown in the accompanying Figure 8 .

\begin{figure}[h]
  \centering
  \includegraphics[width=1\linewidth]{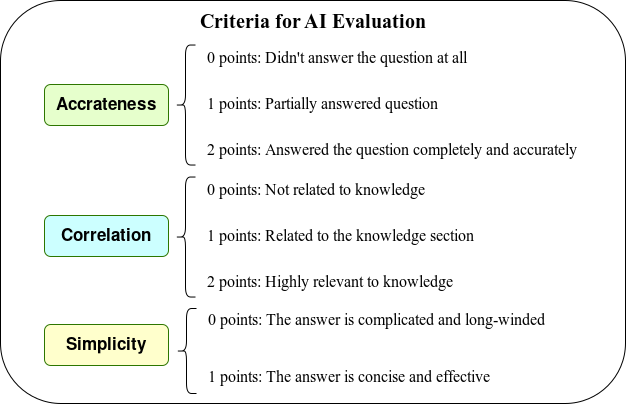}
  \caption{Detail criteria for AI Evaluation}
  \label{fig:example}
\end{figure}
\subsection{Evaluation on LLM for extracting entity}
\hspace{1em}The performance of the LLM1 model for extracting question entities was evaluated using several key metrics, including BLEU, ROUGE, METEOR, BERTScore, and accuracy (Acc), for the research object, spectral method, and question type to be extracted in the input question. At the same time, the performance of the Llama2-7b model after Lora fine-tuning and the baseline Chat-GPT model in extracting entities was compared.
\begin{figure}[h]
  \centering
  \includegraphics[width=1\linewidth]{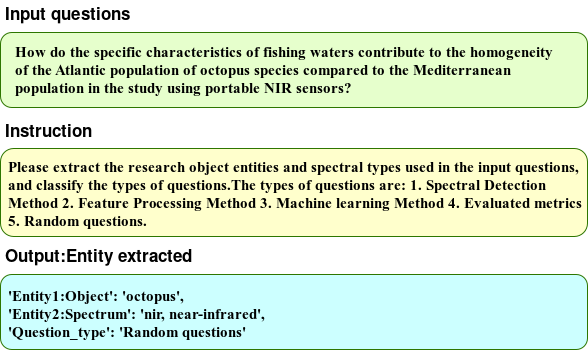}
  \caption{Example for the process of extracting entity}
  \label{fig:example}
\end{figure}

\hspace{1em}LLM1 significantly outperformed the baseline Chat-GPT model in all metrics, and the experimental results are shown in Table 2. For extracting research objects from questions, LLM1 achieved higher scores in BLEU, ROUGE1, METEOR, and BERTScore accuracy. Similarly, for extracting spectral methods mentioned in questions, LLM1 showed excellent performance. LLM1 also had significantly higher accuracy on both tasks.

\hspace{1em}In terms of identifying question types, LLM1 consistently provided better results than Chat-GPT, demonstrating its enhanced ability to understand and extract relevant entities in the context of spectral detection.

\begin{table*}[h!]
\caption{Evaluation results of LLM1 for extracting entity}
\label{tab:Evaluation results of LLM for extracting entity}
\renewcommand{\arraystretch}{1.5}
\centering
\scalebox{0.85}{
\begin{tabular}{lcccccccccccccc}
\hline
\multirow{2}{*}{\textbf{Models}} & \multicolumn{3}{c}{\textbf{Bleu}} & \multicolumn{3}{c}{\textbf{Rouge1}} & \multicolumn{3}{c}{\textbf{meteor}} & \multicolumn{3}{c}{\textbf{Bert\_score\_precision}} & \multicolumn{2}{c}{\textbf{Acc}} \\ \cline{2-15}
                                 & \textbf{Et1} & \textbf{Et2} & \textbf{Q\_type} & \textbf{Et1} & \textbf{Et2} & \textbf{Q\_type} & \textbf{Et1} & \textbf{Et2} & \textbf{Q\_type} & \textbf{Et1} & \textbf{Et2} & \textbf{Q\_type} & \textbf{Et2} & \textbf{Q\_type} \\ \hline
\textbf{Llama2-7b}               & 0.087        & 0.423        & 0               & 0.528        & 0.871        & 0.978           & 0.439        & 0.831        & 0.943           & 0.911        & 0.972        & 0.997           & 0.882        & 0.977           \\
\textbf{Chat-GPT}                & 0.02         & 0.1          & 0               & 0.093        & 0.445        & 0.711           & 0.040        & 0.278        & 0.681           & 0.874        & 0.894        & 0.956           & 0.584        & 0.534           \\ \hline
\end{tabular}}
\end{table*}

\subsection{Evaluation of the knowledge Retrieve}
\hspace{1em}After extracting entities from the input question using LLM1, the retrieval method will retrieve relevant papers based on the entity keywords. The retrieval methods used in the experiment include the Bag of Words model, BM25, and TF-IDF cosine similarity retrieval methods.

\hspace{1em}The Bag of Words (BoW) model is a basic and widely used method in text representation, in which each document is represented as an unordered collection of words without considering grammar and word order, but considering the frequency of words. Although the BoW model is simple, it usually lacks contextual understanding, which leads to low accuracy in complex retrieval tasks. BM25 is a ranking function based on a probabilistic retrieval framework that enhances the traditional TF-IDF method by considering word frequency, document length, and inverse document frequency. This method is recognized to be effective in information retrieval tasks. The TF-IDF (Term Frequency-Inverse Document Frequency) cosine similarity retrieval method combines TF-IDF weighting with cosine similarity to measure the angle between document vectors in the vector space model. This method effectively captures the relevance between documents and queries, thereby improving the accuracy of retrieval tasks.
\begin{table}[!ht]
\caption{Evaluation results of different knowledge retrieve methods}
\label{tab:Evaluation results of different knowledge retrieve methods}
    \renewcommand{\arraystretch}{1.3}
    \centering
    \scalebox{0.95}{
    \begin{tabular}{lc}
    \hline
        Methods & Accuracy(\%) \\ \hline
        Bag-of-Words model & 46.40 \\
        BM25 & 80.00 \\
        TF-IDF cosine similarity retrieval method & 82.80 \\ \hline
    \end{tabular}}
\end{table}

\hspace{1em}The accuracy of each method is calculated as follows: If the papers retrieved using the entity keyword-based retrieval method match the actual papers corresponding to the keyword, the accuracy is 100\%. Then the average accuracy of all queries in the database is calculated as the final accuracy of each method. The experimental results are shown in Table 2.

\hspace{1em}As can be seen from the table, the TF-IDF cosine similarity retrieval method has the highest accuracy of 82.80\%, followed by BM25, which is 80.00\%. The Bag-of-Words model has the lowest accuracy of 46.40\%. These results demonstrate the superior performance of TF-IDF cosine similarity retrieval in accurately retrieving relevant papers based on the extracted entity keywords.

\hspace{1em}By adopting these methods, we are able to compare their retrieval performance and determine that the TF-IDF cosine similarity retrieval method outperforms other methods in terms of accuracy.
\subsection{Evaluation on LLM for generating response}
\hspace{1em}The performance of LLM2 models (particularly Llama3-8b and Llama2-13b) was evaluated using several key metrics, including BLEU, ROUGE, METEOR, BERTScore, and AI evaluation, and the experimental results are shown in Table 4. These results are compared with the baseline Chat-GPT model. Both Llama2-13b and Llama3-8b models were fine-tuned using the LoRA technique.

\begin{figure}[h]
  \centering
  \includegraphics[width=1\linewidth]{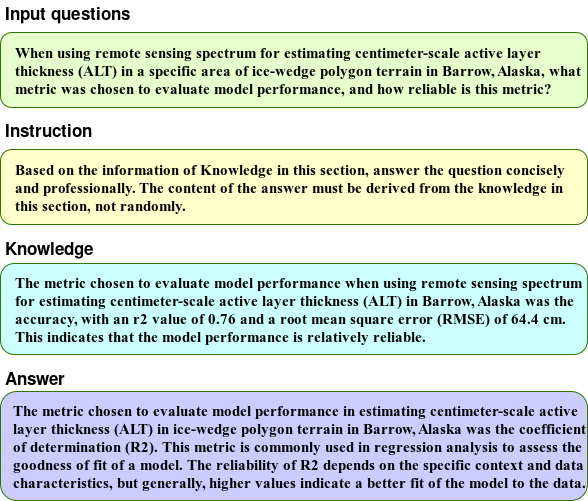}
  \caption{Example for generating response}
  \label{fig:example}
\end{figure}

\hspace{1em}Llama3-8b achieved the highest scores across most metrics, with a BLEU score of 0.304, ROUGE score of 0.558, METEOR score of 0.503, BERTScore of 0.881, and an AI evaluation score of 4.2. Llama2-13b followed with a BLEU score of 0.158, ROUGE score of 0.385, METEOR score of 0.344, BERTScore of 0.861, and an AI evaluation score of 3.5. The baseline Chat-GPT model scored significantly lower across all metrics, with a BLEU score of 0.001, ROUGE score of 0.065, METEOR score of 0.033, BERTScore of 0.834, and an AI evaluation score of 2.8.

\begin{table}[!ht]
\caption{Evaluation results of LLM2 for generating response}
\label{tab:Evaluation results of LLM2 for generating response}
    \renewcommand{\arraystretch}{1.3}
    \centering
    \scalebox{0.85}{
    \begin{tabular}{lccccc}
    \hline
        Models & Bleu & rouge & meteor & Bert\_score & AI evaluate \\ \hline
        Llama3-8b & 0.304 & 0.558 & 0.503 & 0.881 & 4.2 \\ 
        Llama2-13b & 0.158 & 0.385 & 0.344 & 0.861 & 3.5 \\
        Chat-GPT & 0.001 & 0.065 & 0.033 & 0.834 & 2.8 \\ \hline
    \end{tabular}}
\end{table}

\hspace{1em}These results indicate that the Llama3-8b model, fine-tuned with LoRA, performed the best in generating responses, demonstrating superior language generation capabilities and relevance to the spectral detection context compared to both Llama2-13b and the baseline Chat-GPT model.

\section{Conclusion}
\hspace{1em}In this paper, to enable researchers in the field of spectral detection to retrieve spectral related knowledge faster and more accurately, we designed a fast and reliable spectral detection question answering system based on the SDAAP dataset and a large language model. Considering that professional literature is a valuable treasure trove of advanced knowledge, research results, and engineering methods, we first proposed the Spectral Detection and Analysis Paper (SDAAP) dataset as a basic knowledge source. Subsequently, we developed an automatic question answering framework in related fields based on SDAAP, which uses the Llama2-7b model fine-tuned by LoRA to parse and extract entities and question formats present in the query, and uses the parsed results as query parameters to retrieve relevant spectral detection knowledge through the cosine similarity retrieval method. Finally, the Llama3-8b fine-tuned by LoRA can refer to these retrieved knowledge to generate responses to input queries. Through experiments on the spectral detection knowledge question answering dataset we proposed, the two fine-tuned large language models used to extract entities and generate answers achieved higher accuracy and reliability in generating responses. Both models performed well in commonly used natural language indicator evaluations, with evaluation scores far exceeding Chat-GPT. They can accurately and quickly extract entities and generate answers, respectively, highlighting the domain adaptation potential of LLM in professional fields.

\section{Limitation}
\textbullet Datasets: A major limitation of this study is the limited number of question-answer pairs and the reliance on a limited amount of literature data, which may inhibit the applicability and validity of the model. On the other hand, the manual cleaning process applied to the data could still be problematic. To address this issue, future research could focus on expanding the task to generate more question-answer pairs; or further structuring the relevant literature database and constructing knowledge graphs to enhance the interconnectivity of the IFT dataset and improve the performance of the model in the vertical domain.

\textbullet Evaluation methodology: In addition to the Type 1 questions, we select metrics such as Bleu and Rouge to measure the similarity between the model output and the golden facts. However, these metrics may not be well suited for model evaluation in verticals, and relying solely on the similarity of the output to the golden facts may not be a valid reflection of the quality of the generated answers. It is worth noting that in the first category of questions, metrics such as Bleu still yield high scores even if the answers are wrong on key spectral species. To address this issue, in the future we will consider introducing entirely new evaluation metrics or using expert evaluation for some of the responses.


\end{document}